\documentclass[
reprint,
amsmath,
amssymb,
aps,
]{revtex4-2}

\usepackage{graphicx} 
\usepackage{dcolumn}  
\usepackage{bm}       
\usepackage{hyperref} 
\usepackage{calc}
\usepackage{physics}
\usepackage{bbm}
\usepackage{verbatim}

\begin{document}

\preprint{APS/123-QED}

\title{Practical verification protocols for analog quantum simulators}

\author{Ryan Shaffer}
\email{ryan.shaffer@berkeley.edu}
\author{Eli Megidish}
\author{Joseph Broz}
\author{Wei-Ting Chen}
\author{Hartmut H\"{a}ffner}
\affiliation{
 Department of Physics, University of California, Berkeley, California 94720, USA
}

\date{\today}

\begin{abstract}
Analog quantum simulation is expected to be a significant application of near-term quantum devices. Verification of these devices without comparison to known simulation results will be an important task as the system size grows beyond the regime that can be simulated classically.
We introduce a set of experimentally-motivated verification protocols for analog quantum simulators, discussing their sensitivity to a variety of error sources and their scalability to larger system sizes.
We demonstrate these protocols experimentally using a two-qubit trapped-ion analog quantum simulator and numerically using models of up to five qubits.
\end{abstract}

\maketitle

%
%
\section{\label{sec:introduction}Introduction}

%
%
Quantum simulation has long been proposed as a primary application of quantum information processing \cite{Feynman1982SimulatingComputers}.
In particular, analog quantum simulation, in which the Hamiltonian evolution of a particular quantum system is directly implemented in an experimental device, 
is projected to be an important application of near-term quantum devices \cite{Preskill2018QuantumBeyond}, with the goal of providing solutions to problems that are infeasible for any classical computer in existence.
Because the obtained solutions to these problems cannot always be checked against known results, a key requirement for these devices will be the ability to verify that the desired interactions are being carried out faithfully \cite{Cirac2012GoalsSimulation, Hauke2012CanSimulators, Johnson2014WhatSimulator}. 
If a trusted analog quantum simulator is available, then one can certify the behavior of an untrusted analog quantum simulator \cite{Wiebe2014HamiltonianResources}.
But in the absence of a trusted device, provable verification is essentially intractable for systems of interest that are too large to simulate classically \cite{Cirac2012GoalsSimulation}.
Therefore, in the near-term, we see a need to develop pragmatic techniques to verify these devices and thus increase confidence in the results obtained.

Many experimental platforms have been used to perform analog quantum simulations of varying types, including devices based on
neutral atoms \cite{Bloch2012QuantumGases,Bernien2017ProbingSimulator,Gross2017QuantumLattices},
trapped ions \cite{Blatt2012QuantumIons,Gorman2018EngineeringSimulator,Kokail2019Self-verifyingModels},
photons \cite{Aspuru-Guzik2012PhotonicSimulators},
and superconducting circuits \cite{Houck2012On-chipCircuits}.
In such works, validation of simulation results is typically performed by comparison to results calculated analytically or numerically in the regime where such calculation is possible.
In addition, a technique for self-verification has been proposed and demonstrated \cite{Kokail2019Self-verifyingModels} which measures the variance of the energy to confirm that the system has reached an eigenstate of the Hamiltonian.
However, this technique does not verify whether the desired Hamiltonian has been implemented faithfully.

One method which has been proposed for analog simulation verification is to run the dynamics forward and backward for equal amounts of time \cite{Cirac2012GoalsSimulation}, commonly known as a Loschmidt echo \cite{Jalabert2001Environment-independentSystems,Gorin2006DynamicsDecay}, which ideally returns the system to its initial state. Such a method is not able to provide confidence that the parameters of the simulation are correct, nor can it detect some common sources of experimental error such as slow environmental fluctuations or crosstalk between various regions of the physical device. However, it is naturally scalable and is straightforward to implement experimentally, provided that a time-reversed version of the analog simulation can be implemented. An extension of this method similar to randomized benchmarking has also been proposed \cite{Derbyshire2020RandomizedSetting}, although this suffers from the same shortcomings just mentioned.

Another natural candidate for verification of analog simulations is to build multiple devices capable of running the same simulation and to compare the results across devices, which is a technique that has been demonstrated for both gate-based devices \cite{Greganti2019Cross-verificationDevices} and analog simulators \cite{Elben2020Cross-PlatformDevices}. This technique has the obvious difficulty of requiring access to additional hardware, in addition to the fact that it may be difficult to perform the same analog simulation across multiple types of experimental platforms.

Experimentalists building analog quantum simulators are in need of practical proposals for validating the performance of these devices. 
Ideally such a protocol can be executed on a single device, can provide confidence that the target Hamiltonian is correctly implemented, and can be scaled to large systems.
In this work, we aim to address these goals by introducing a set of experimentally practical approaches to the task of validating the performance of analog quantum simulators.


\section{\label{sec:results}Results}

%
%
\subsection{\label{sec:verification-protocols}Overview of verification protocols}


The task of analog quantum simulation involves configuring a quantum system in some initial state, allowing it to evolve according to some target Hamiltonian for a particular time duration, and then analyzing one or more observables of interest. A verification protocol for this process should provide some measure of how faithfully the device implements the target Hamiltonian.

We claim that a useful protocol for verification
of analog quantum simulators should have the following attributes:

\textit{Independent of numerical calculations of the system dynamics.}
We should not need to rely on comparison of the
analog simulation results to numerically-calculated dynamics of the full system,
since simulations of interest will be performed in regimes where
numerical calculation is infeasible.

\textit{Efficient to measure.}
Verification protocols should leave the system in or near a basis state, rather than in some arbitrary state. This allows characterization of the final state by making only a small number of measurements. This allows us to circumvent the need for more intensive procedures such as full state tomography, which in turn reduces the experimental overhead.

\textit{Sensitive to many experimental error
sources.} The main objective of a verification
protocol is to measure experimental imperfections. If a
protocol is not sensitive to some potential sources of
experimental error in the simulation, it cannot give us maximal
confidence in the results.

\textit{Applicable to near-term analog quantum simulators.}
Unlike many benchmarking protocols for digital, gate-based quantum computers, we are not seeking a protocol which can
give fine-grained information about the
fidelity of a particular operation, but rather an approach which can give us coarse-grained information about
the reliability of a noisy simulation.

\textit{Scalable to large systems.} Many interesting
near-term analog quantum simulations will likely be performed in regimes
where the system size is relatively large (many tens or
hundreds of qubits). A useful verification protocol for
such devices should be efficiently scalable to these
system sizes, given reasonable assumptions.

%
%
\begin{table*}
\caption{\label{tab:protocol-summary}Summary of characteristic error sensitivity, hardware requirements, and scalability limits for proposed verification protocols for analog quantum simulators.}
\begin{ruledtabular}
\begin{tabular}{>{\raggedright}p{0.13\linewidth} >{\raggedright}p{0.24\linewidth} >{\raggedright}p{0.26\linewidth} >{\raggedright}p{0.32\linewidth}}
\textbf{Protocol} & \textbf{Error sensitivity} & \textbf{Hardware requirements} & \textbf{Scalability limits}
\tabularnewline
\hline
Time-reversal analog verification & Fast incoherent noise & Implement time-reversed analog simulation & None inherent
\tabularnewline
\hline
Multi-basis analog verification & Fast incoherent noise, shot-to-shot parameter fluctuation & Implement time-reversed analog simulation in alternate basis and single-qubit rotations & None inherent
\tabularnewline
\hline
Randomized analog verification & Fast incoherent noise, shot-to-shot parameter fluctuation, parameter miscalibration,
crosstalk & Implement time-reversed analog simulation and ability to turn Hamiltonian terms on/off individually & Approximate inverse compilation procedure requires simulation of dynamics; protocol must be performed on subsets of larger systems
\end{tabular}
\end{ruledtabular}
\end{table*}

%
%
\begin{figure}

\includegraphics[width=0.49\linewidth]{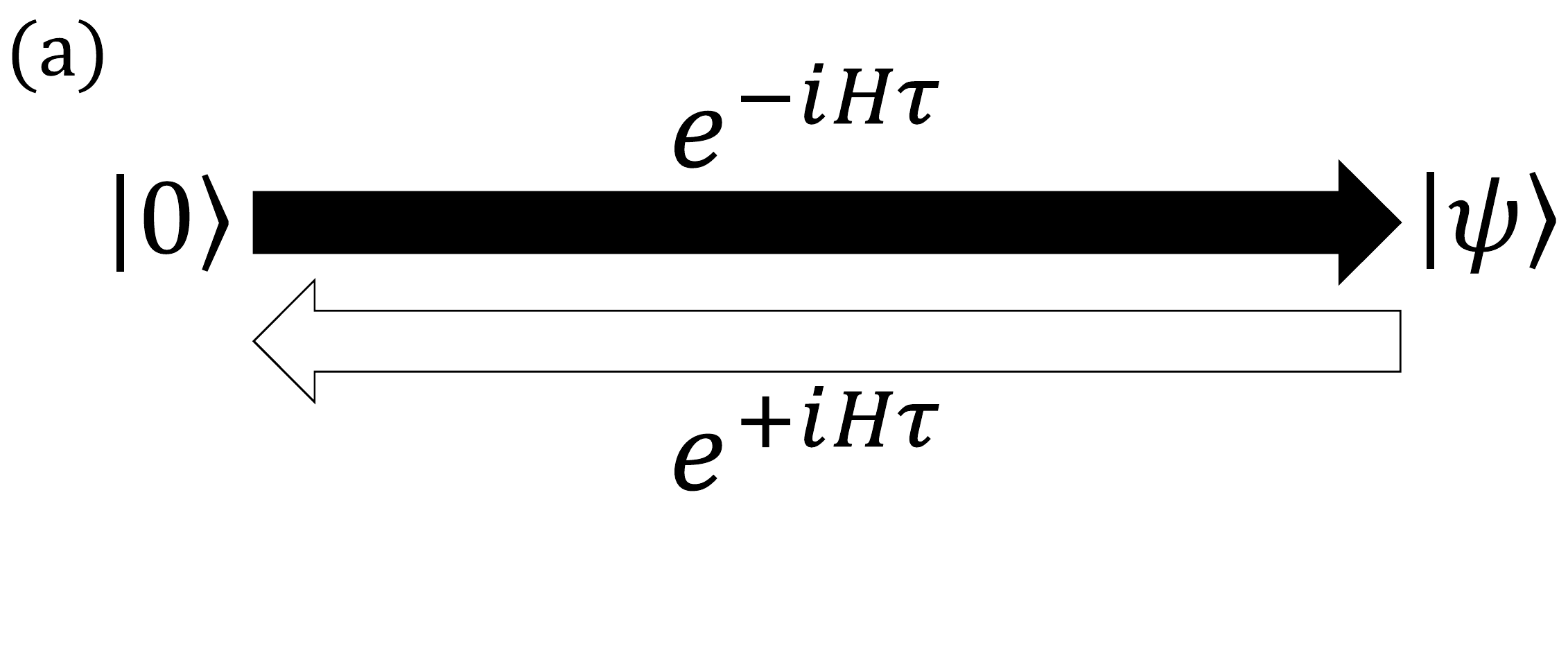}
\includegraphics[width=0.49\linewidth]{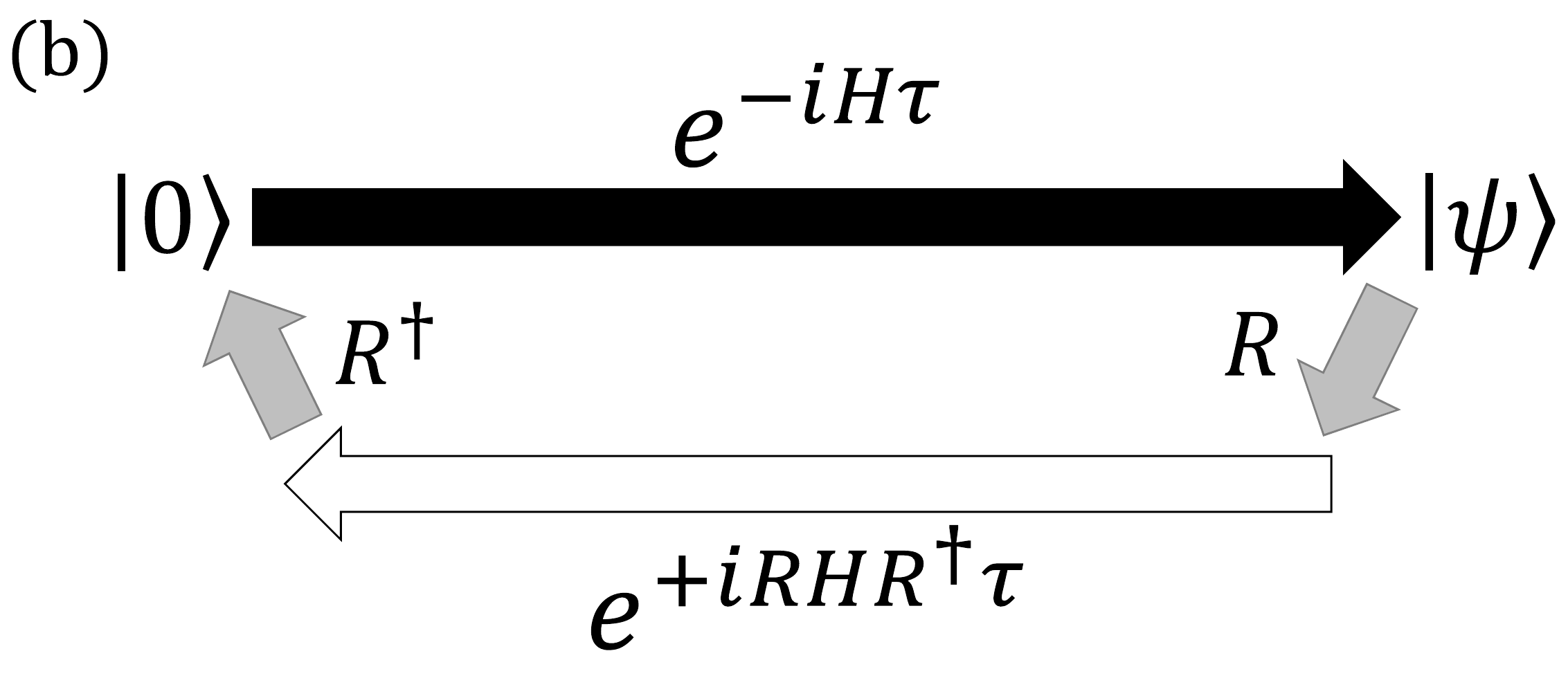}
\includegraphics[width=0.63\linewidth]{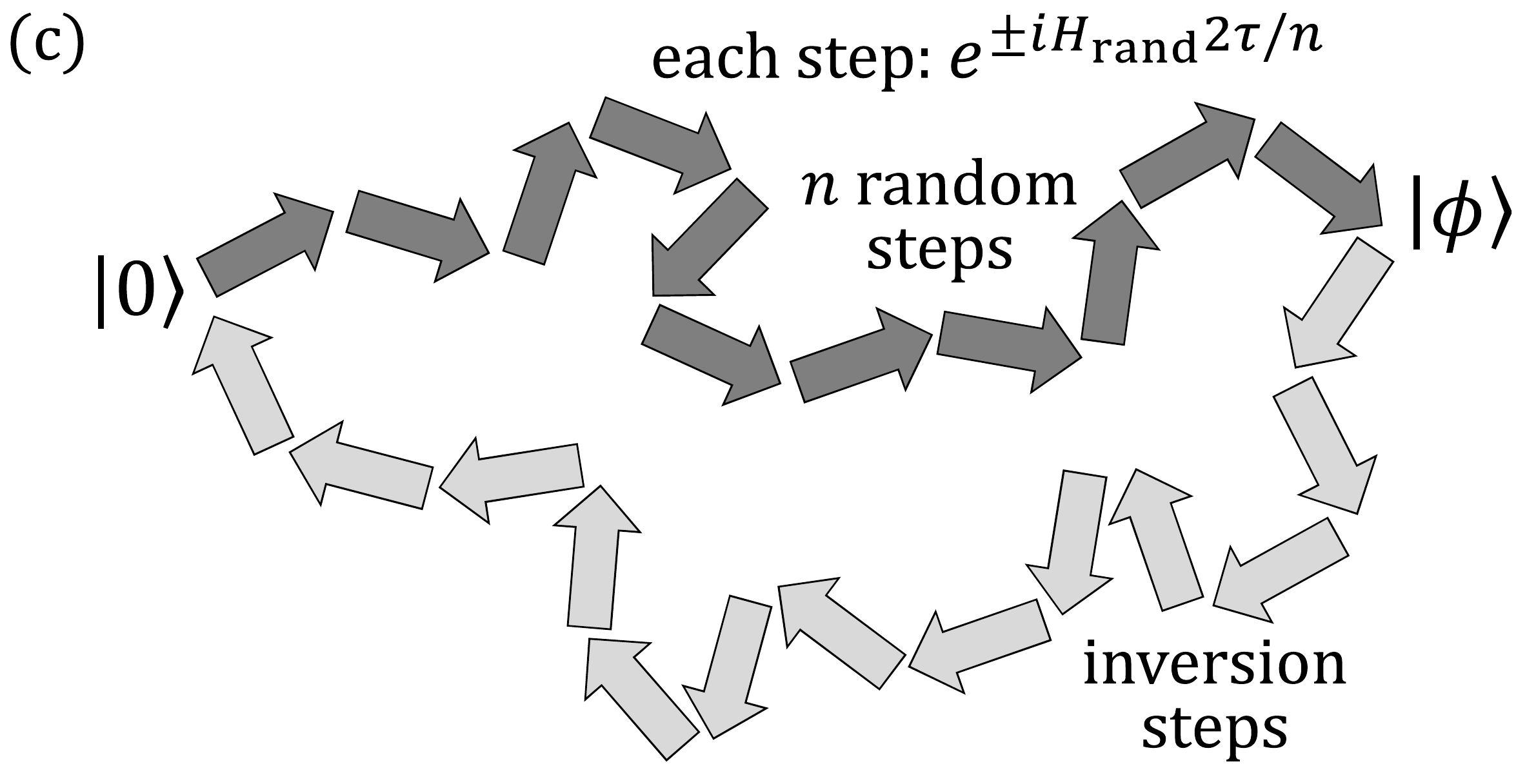}
\caption{\label{fig:sequence-illustrations}\textbf{Illustration of verification protocols for analog quantum simulators.} Various protocols yield information about the accuracy of a quantum simulator by propagating a state along a closed loop and verifying to what degree the system returns to its original state, labeled here as $\ket{0}$.
The state $\ket{\psi}$ denotes the state of the system after applying the dynamics of Hamiltonian $H$ for a time $\tau$, whereas the state $\ket{\phi}$ denotes an arbitrary state. \\
(a) \textit{Time-reversal analog verification:} Running an analog simulation forward in time, followed by the same analog simulation backward in time. \\
(b) \textit{Multi-basis analog verification:} Running an analog simulation forward in time, rotating the state, performing the backward simulation by an analog version in the rotated basis, and finally rotating the state back. \\
(c) \textit{Randomized analog verification:} Running a random sequence of subsets of the Hamiltonian terms (denoted as $H_{\textrm{rand}}$), followed by an inversion sequence of subsets of the Hamiltonian terms which has been calculated to return the system approximately to a basis state.
}
\end{figure}



In this work, we propose a set of three verification protocols for analog quantum simulators which exhibit many of these attributes. These are illustrated in Fig.~\ref{fig:sequence-illustrations}.
The overarching strategy for each protocol, inspired by the Loschmidt echo procedure, involves asking the simulator to evolve a system from some known initial state through a closed loop in state space, eventually returning to its initial state. By using a basis state as the initial (and final) state, we can efficiently measure the success of this procedure. A number of strategies exist to construct such a closed loop, with varying pros and cons. We use a few of these strategies to construct the proposed verification protocols. These protocols are summarized in Table~\ref{tab:protocol-summary}, including some types of experimental noise to which each protocol is sensitive, the hardware requirements for implementing each protocol, and the scalability constraints of each protocol.

First,
we propose a time-reversal analog verification protocol, in which the simulation is run both forward and backward in time. As illustrated in Fig.~\ref{fig:sequence-illustrations}(a), this approach simply performs a Loschmidt echo to reverse the time dynamics of the simulation and then verifies that the system has returned to its initial state. However, because the system traverses the same path in state space in the forward and backward directions, it is insensitive to many types of experimental errors, including systematic errors such as miscalibrations in the Hamiltonian parameters or crosstalk between sites.

To increase the susceptibility to systematic errors,
we propose a multi-basis analog verification protocol, as shown in Fig.~\ref{fig:sequence-illustrations}(b). This is a variant of the time-reversal protocol in which a global rotation is performed on the system after the completion of the forward evolution, and the backward evolution is then performed in the rotated basis. Because this requires a physical implementation of the analog simulation in an additional basis, it will provide sensitivity to any systematic errors that differ between the two bases. For example, errors due to some types of shot-to-shot noise may be enhanced and not cancel out as in the previous protocol.

However, we note that the previous two protocols may still be insensitive to many types of errors, such as miscalibration or the presence of unwanted constant interaction terms. To address this,
we introduce a randomized analog verification protocol, which consists of running randomized analog sequences of subsets of the target Hamiltonian terms, as depicted in Fig.~\ref{fig:sequence-illustrations}(c).
In particular, we choose a set of unitary operators consisting of short, discrete time steps of each of the terms of the Hamiltonian to be simulated, which may be in either the forward or backward direction.
We randomly generate long sequences of interactions, each consisting of a subset of these unitary operators, which evolves the system to some arbitrary state.
We then use a Markov chain Monte Carlo search technique to approximately compile an inversion sequence using the same set of unitary operators, such that after the completion of the sequence, the system is measured to be in a basis state with high probability.
This scheme is an adaptation of traditional gate-based randomized benchmarking techniques \cite{Emerson2005ScalableOperators, Knill2008RandomizedGates} for use in characterizing an analog quantum simulator.
A key difference is that for a general set of Hamiltonian terms, finding a non-trivial exact inversion of a random sequence is difficult, which is why we instead find an approximate inversion sequence. In principle, this approximation is a limitation on the precision with which this protocol can be used to verify device performance.
However, in practice, the search technique can be used to produce inversion sequences that return a large percentage (e.g., 99\% or more) of the population to a particular basis state, which is enough for the protocol to be useful on noisy near-term devices, since even the most accurate analog quantum simulations typically have fidelities that decay far below this level \cite{Lysne2020SmallSimulations}.


Each verification protocol can then be executed for varying lengths of time, and the measurement results will provide the success probability of each protocol as a function of time. For a system that implements the target Hamiltonian perfectly, one expects this probability to remain constant, with a small offset from unity due to state preparation and measurement errors, as well as the approximation error for the inversion sequence in the randomized protocol. But if the system dynamics are not perfect, one expects the success probability to decrease as a function of time.

For standard randomized benchmarking protocols, the shape of the decay curve provides additional information about the errors, for example, allowing one to distinguish whether the dominant error source affecting the dynamics is Markovian or non-Markovian.
For typical incoherent noise, one expects this to be an exponential decay, but for noise that is non-Markovian \cite{Epstein2014InvestigatingProtocols,Wallman2018RandomizedNoise} or low-frequency \cite{Fogarty2015NonexponentialNoise}, the decay curve may be non-exponential.

However, in general, we make no strong claim about the shape of the decay curves resulting from the analog verification protocols. In particular, randomized benchmarking requires that the gate set must form an $\epsilon$-approximate 2-design, which is true not only of the Clifford group but also of any universal gate set, given that the randomly generated sequences are long enough \cite{Harrow2009Random2-designs}. However, the time-evolution operator generated by a fixed Hamiltonian cannot approach a 2-design without adding a disorder term \cite{Derbyshire2020RandomizedSetting}, which means that we cannot directly apply randomized benchmarking theory for the time-reversal or multi-basis analog verification protocols. And even the randomized analog verification protocol, which is conceptually more similar to randomized benchmarking, does not require that the Hamiltonian terms actually generate a universal gate set or that the generated sequences are long enough to approximate a unitary 2-design.

Nonetheless, the decay curves still contain potentially useful information about the reliability of the analog quantum simulator. The protocols could be used as a tool to assist in calibrating an analog simulation by attempting to minimize the decay. Also, since each protocol has different sensitivities to errors, comparing decay curves from the various protocols may give clues to an experimentalist about the types of errors that are present.


In this work, we treat noise sources in an analog quantum simulation as modifications of the target Hamiltonian. Physically, these could be caused by variations in quantities such as laser intensity, microwave intensity, magnetic fields, or other terms which could create undesired interactions with the system. We can then represent the full Hamiltonian implemented by the system as
\begin{equation}
    \tilde{H}(t) = H + \delta H(t),
\end{equation}
where $H$ is the target Hamiltonian to be simulated, which we assume is time-independent, and
\begin{equation}
    \delta H(t) = \sum_k \lambda_k(t) \,\delta H_k
\end{equation}
represents any unwanted time-dependence and other miscalibrations present in the physical system. We assume that each $\lambda_k(t)$ varies on some characteristic timescale $t_k$. For example, if $\lambda_k(t)$ is a stationary Gaussian process, then $t_k$ may be the decay time of the autocorrelation function
$R(t)= \expval{\lambda_k(0) \lambda_k(t)}$.
We note that there are several distinct regimes:

\textit{Miscalibrations.} $t_k \gg N \tau$, where $N$ is the number of repetitions performed in a quantum simulation experiment, and $\tau$ is the total runtime of each repetition. This regime corresponds to miscalibrations, unwanted interactions, and other noise that varies on a very slow timescale.

\textit{Slow noise.} $N \tau > t_k > \tau$. This corresponds to noise that causes fluctuations from one run of the experiment to the next, but is roughly constant over the course of a single experiment, i.e., shot-to-shot noise.

\textit{Fast noise.} $t_k \ll \tau$. This is the type of fluctuation that is most commonly referred to as ``noise'', i.e., fluctuations in parameters that are much faster than the timescale of a single experiment.

We design verification protocols to detect different subsets of these noise types: the time-reversal analog verification protocol for detecting fast noise, the multi-basis analog verification protocol for additionally detecting some types of slow noise, and finally the randomized analog verification protocol for detecting miscalibrations and other unwanted interactions. These protocols are described and demonstrated in the remainder of this work.

%
%
\subsection{\label{sec:protocoldetails-timereversal}Time-reversal verification protocol}


The time-reversal analog verification protocol consists of the following steps, repeated for various values of $\tau$, which should range over the characteristic time scale of the simulation to be tested:

\textit{Step 1.} Initialize the system state to an arbitrarily-chosen basis state $\ket{i}$.

\textit{Step 2.} Apply the analog simulation for time $\tau$, that is, apply the unitary operator $e^{-i H \tau}$, which ideally takes the system to the state $\ket{\psi}$. (We use the convention $\hbar = 1$ here and throughout this work.)

\textit{Step 3.} Apply the analog simulation with reversed time dynamics for time $\tau$, that is, apply the operator $e^{+i H \tau}$, which ideally takes the system to the state $\ket{i}$.

\textit{Step 4.} Measure the final state in the computational basis. Record the probability that the final state is measured to be $\ket{i}$.

After repeating these steps for various values of $\tau$, a decay curve can be plotted which indicates the success probability of finding the system in the desired state as a function of simulation time.


We first note that this protocol does not provide validation of the values of any time-independent Hamiltonian parameters, because if $\tilde{H}$ is time-independent, $e^{i \tilde{H} \tau} e^{-i \tilde{H} \tau} = \mathbbm{1}$ regardless of whether $\tilde{H}$ is actually the desired Hamiltonian. It does, however, provide sensitivity to fast, incoherent noise that affects the system on a timescale shorter than the simulation time, and it also will detect imperfections in the implementation of the time-reversal itself.

More formally, the forward time-evolution operator from time 0 to $\tau$ can then be written explicitly in terms of a Dyson series as
\begin{equation}
    U_\textrm{fwd} (0, \tau) = \mathcal{T} e^{-i \int_0^\tau dt\,(H+\delta H(t)) },
\end{equation}
where $\mathcal{T}$ is the time-ordering operator. The reverse time-evolution operator from time $\tau$ to $2\tau$ is then
\begin{equation}
    U_\textrm{rev} (\tau, 2\tau) = \mathcal{T} e^{+i \int_\tau^{2\tau} dt\,(H+\delta H(t)) }.
\end{equation}

It is apparent that if the noise terms in the Hamiltonian are constant between times $0$ and $2\tau$, i.e., if $\delta H(t) = \delta H$, then we have
\begin{equation}
    U_\textrm{rev}(\tau, 2\tau)\, U_\textrm{fwd}(0, \tau)
    = e^{+i \tau\,(H+\delta H) }\, e^{-i \tau\,(H+\delta H) } = \mathbbm{1}
\end{equation}
and thus applying the forward and reverse time-evolution operators will return the system to its initial state.

However, this is not true in general if the noise terms have time-dependence. We can illustrate this by making a simplifying assumption that the noise is piecewise constant between times $0$ and $2\tau$ as
\begin{equation}
    \delta H(t) =
    \begin{cases}
        \delta H_1 & 0 \le t < \tau \\
        \delta H_2 & \tau \le t < 2 \tau
    \end{cases}
\end{equation}
where $\delta H_1$ and $\delta H_2$ are non-commuting in general. We then perform a first-order Baker-Campbell-Hausdorff approximation, which shows that
\begin{align}
    U_\textrm{rev}(\tau, 2\tau)\, &U_\textrm{fwd}(0, \tau) \nonumber \\
    &= e^{+i \tau\,(H+\delta H_2) }\, e^{-i \tau\,(H+\delta H_1) } \\
    &\approx e^{+i \tau (\delta H_2 - \delta H_1 + [H+\delta H_1,\, H+\delta H_2]/2)} .
\end{align}
In the general case where $\delta H_1 \neq \delta H_2$, this quantity will not be equal to the identity. A similar argument also holds if the noise terms vary on faster timescales. That is, if $\delta H(t)$ contains one or more noise terms such that $\lambda_k(t)$ has a correlation time $t_k \ll \tau$, then the product of the forward and reverse time-evolution operators will not be equal to the identity in general, and the system will not return to its initial state.


The time-reversal analog verification protocol requires only that the analog quantum simulator is capable of implementing the time-reversed dynamics of the desired simulation, that is, the signs of each of the Hamiltonian terms can be negated.
Because there are no numerical calculations required, the protocol is independent of the size of the system, and its scalability has no inherent limitations, outside of any physical limitations involved in implementing the analog simulation itself in both directions.

%
%
\subsection{\label{sec:protocoldetails-multibasisanalog}Multi-basis analog verification protocol}


The multi-basis analog verification protocol consists of the following steps, repeated for various values of $\tau$, which should range over the characteristic time scale of the simulation to be tested:

\textit{Step 1.} Initialize the system state to an arbitrarily-chosen basis state $\ket{i}$.

\textit{Step 2.} Apply the analog simulation for time $\tau$, that is, apply the unitary operator $e^{-i H \tau}$, which ideally takes the system to the state $\ket{\psi}$.

\textit{Step 3.} 
Apply a basis transformation $R$ to the system to take it to the state $R \ket{\psi}$, with $R$ chosen such that both $R$ and the rotated inverse Hamiltonian
\begin{equation}
\label{eq:hprime-rotated-basis}
    H' = R H R^\dag
\end{equation}
are implementable. For example, if the target Hamiltonian is 
\begin{equation}
    H = \sigma_x^{(1)} \sigma_x^{(2)},
\end{equation}
one could choose 
\begin{equation}
    R = \sqrt{\sigma_y}^{(1)} + \sqrt{\sigma_y}^{(2)}
\end{equation}
if and only if the analog quantum simulator can physically implement the interactions $R$, $H$, and 
\begin{equation}
    H' = R H R^\dag = \sigma_z^{(1)} \sigma_z^{(2)}.
\end{equation}

\textit{Step 4.} Apply the analog simulation in the rotated basis and with reversed time dynamics for time $\tau$, that is, apply the operator $e^{+i H' \tau}$, which ideally takes the system to the state $R \ket{i}$.

\textit{Step 5.} Apply the inverse of the rotation performed in Step 3, that is, apply a global $-\pi/2$ rotation $R^\dag$ to the system, which ideally takes the system back to the initial state $\ket{i}$.

\textit{Step 6.} Measure the final state in the computational basis. Record the probability that the final state is measured to be $\ket{i}$.

After repeating these steps for various values of $\tau$, a decay curve can be plotted which indicates the success probability of finding the system in the desired state as a function of simulation time.


We note that this protocol will detect errors such as miscalibrations or slow fluctuations if the strength of these errors differs in the two bases.
Specifically, if $\tilde{H}$ and $\tilde{H}'$ are the implementations in the two bases which contain noise terms $\delta H(t)$ and $\delta H'(t)$, respectively, then the forward and reverse time-evolution operators can be written as
\begin{align}
    U_\textrm{fwd} (0, \tau) &= \mathcal{T} e^{-i \int_0^\tau dt\,(H+\delta H(t)) }, \\
    U_\textrm{rev} (\tau, 2\tau) &= \mathcal{T} e^{+i \int_\tau^{2\tau} dt\,(H'+\delta H'(t)) }.
\end{align}
Then, even in the simplest case where we have time-independent noise terms $\delta H(t) = \delta H$ and $\delta H'(t) = \delta H'$, we see that applying the forward and reverse time-evolution operators and the appropriate basis-change operators $R$ and $R^\dag$, gives
\begin{align}
    R^\dag\, U_\textrm{rev}(\tau, 2\tau)\,&R \, U_\textrm{fwd}(0, \tau) \nonumber \\ 
    &= R^\dag\, e^{+i \tau\,(H'+ \delta H') } R \, e^{-i \tau\,(H+ \delta H) } \\
    &\approx e^{+i \tau (\delta H'' - \delta H + [H+\delta H,\, H+\delta H'']/2)} ,
\end{align}
where we have defined 
\begin{equation}
    \delta H'' = R^\dag\, \delta H'\, R
\end{equation}
as the rotation of $\delta H'$ into the original basis, and where we use the fact from Eq.~(\ref{eq:hprime-rotated-basis}) that $R^\dag H' R = H$. We assume here for simplicity that $R$ and $R^\dag$ are implemented ideally.

We observe again that the resulting quantity is not equal to the identity in the general case where $\delta H \neq \delta H''$, as well as in the cases where $\delta H$ and $\delta H''$ are non-commuting with each other or with $H$. So we can conclude that in the case that the noise terms $\delta H(t)$ and $\delta H'(t)$ vary independently of each other, even if their correlation times are much longer than the timescale of a single experiment, the system will not return to its initial state when these time-evolution operators are applied.


The multi-basis analog verification protocol requires that the analog quantum simulator implements the desired Hamiltonian in at least two separate bases. For example, a trapped-ion quantum simulator may implement a nearest-neighbor coupling term using both a $\sigma_x \sigma_x$ M{\o}lmer-S{\o}rensen interaction \cite{Srensen1999QuantumMotion} and a $\sigma_z \sigma_z$ geometric phase gate interaction \cite{Duan2001GeometricComputation}, which are equivalent up to a basis change. Likewise, a simulator based on superconducting qubits could implement entangling interactions in multiple bases, for example, b\textsc{swap} interactions using different phases of the microwave drive \cite{Poletto2012EntanglementExcitation}.
(Alternatively, if the device cannot implement the analog simulation in a different basis, but does implement a full universal gate set for quantum computation, the Hamiltonian may be implemented in a digital manner in an alternate basis via Trotterization.) 

In addition to the multi-basis requirement, the device must also have the ability to perform single-qubit rotations in order to make the necessary basis change.
But there are no numerical calculations required in advance, and thus the protocol itself is independent of the size of the system and has no inherent scalability limitations, outside of any limitations in performing the actual analog simulation in the two necessary bases.

%
%
\subsection{\label{sec:protocoldetails-randomizedanalog}Randomized analog verification protocol}


It turns out that the previous two protocols cannot detect all types of errors. Most notably, neither protocol verifies that the simulation actually implements the target Hamiltonian $H$. Errors due to parameter miscalibration or the presence of unwanted constant interaction terms would not be detectable using these schemes.

To address this, we introduce a third protocol,
which consists of running randomized analog sequences of subsets of the target Hamiltonian terms. In particular, we choose a set of unitary operators consisting of short, discrete time steps of each of the terms of the Hamiltonian to be simulated.
We randomly generate long sequences of interactions, each consisting of a subset of these unitary operators, which evolves the system to some arbitrary state.
We then use a stochastic search technique to approximately compile the inverse of these sequences using the same set of unitary operators, which produces another sequence of interactions.  When appended to the original sequence the system returns to the initial state (or another basis state) with high probability.

This protocol is inspired by randomized benchmarking (RB) protocols, which are often used for characterization of gate-based devices \cite{Emerson2005ScalableOperators, Knill2008RandomizedGates, Magesan2011ScalableProcesses, Magesan2012EfficientBenchmarking, Magesan2012CharacterizingBenchmarking, Gambetta2012CharacterizationBenchmarking, Gaebler2012RandomizedGates}.
Most commonly, RB involves generating many random sequences of Clifford gates and appending to each sequence an inversion Clifford.
Ideally, in the absence of errors, the execution of each sequence should return all of the population to a well-known basis state. Measuring the actual population of the desired basis state after the execution of each sequence allows one to calculate a metric related to the average gate fidelity of the device, which can be used to compare the performance of a wide variety of physical devices.

We note that traditional RB has limited scalability due to the complexity of implementing multi-qubit Clifford gates, and has been demonstrated only for up to three qubits \cite{McKay2019Three-QubitBenchmarking}; however, RB-like protocols have been demonstrated on larger systems \cite{Proctor2019DirectDevices, Erhard2019CharacterizingBenchmarking}.

\begin{figure}

\includegraphics[width=0.99\linewidth]{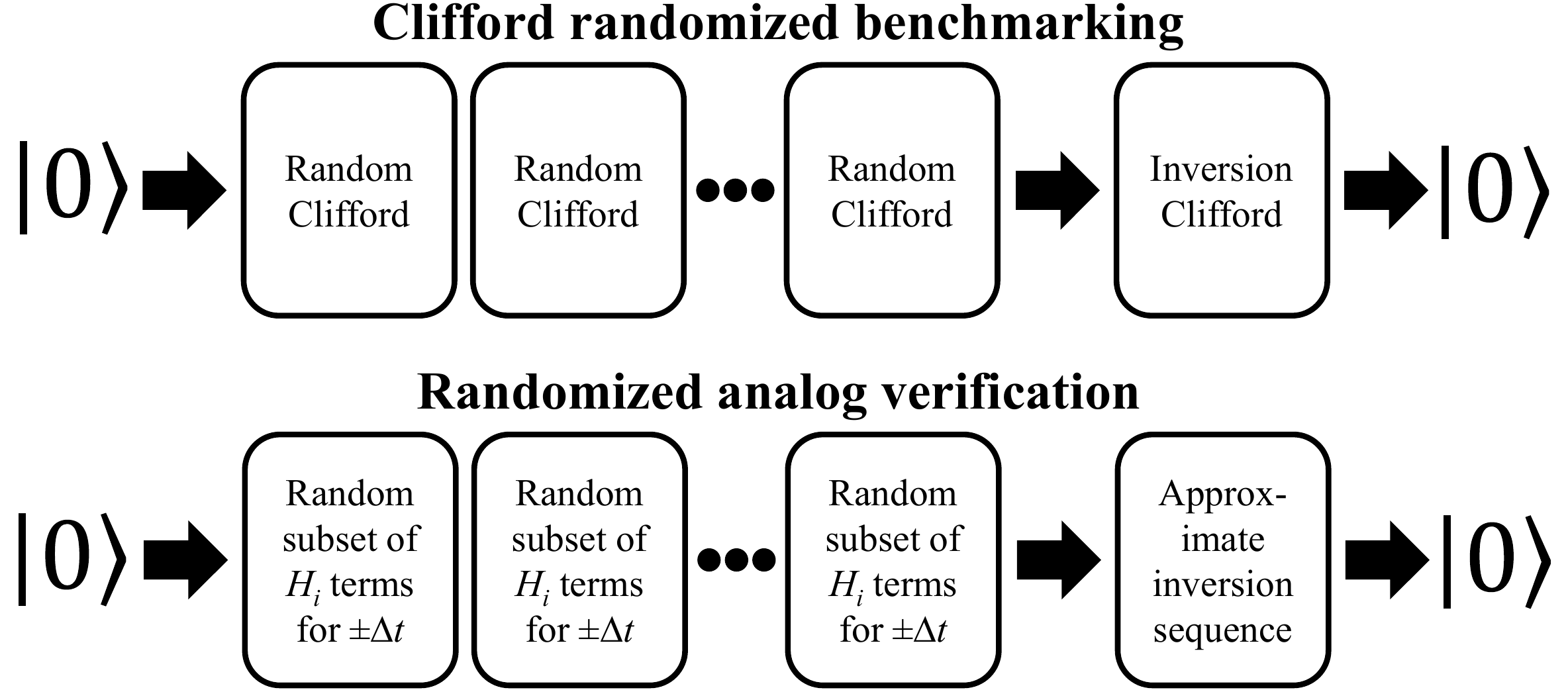}

\caption{\label{fig:randomizedanalog-overview}
    \textbf{High-level comparison of traditional randomized benchmarking and the randomized analog verification protocol.}
    Both protocols involve generating a sequence that starts and ends in a known basis state, which is denoted $\ket{0}$ in this figure for simplicity, and proceed by simply making a series of random choices. For traditional RB, the inversion Clifford is calculated deterministically based on the preceding sequence of random Cliffords. For randomized analog verification, the inversion sequence is compiled approximately via a stochastic search procedure.
}

\end{figure}

Fig.~\ref{fig:randomizedanalog-overview} contains an illustration comparing the randomized analog verification protocol with the traditional Clifford-based RB protocol. We note that this protocol significantly differs from a recently-proposed technique for benchmarking analog devices \cite{Derbyshire2020RandomizedSetting} in that we construct the approximate inversion sequence independently of the initial randomly-generated sequence, which in general prevents miscalibrations and constant errors from cancelling out during the inversion step. We also implement the protocol using subsets of the Hamiltonian terms, which lends itself to scalability.

We write the target Hamiltonian as a sum of terms
\begin{equation}
    H = \sum_{i=1}^{m} H_i,
\end{equation}
where we assume that the simulator can enable both the forward and time-reversed version of each $H_i$ independently of the others. We note that this protocol, in addition to being sensitive to implementation errors in the time-reversal, will also be affected by experimental errors in the enabling or disabling of the individual Hamiltonian terms.

We then repeat the following steps for various values of $\tau$, which is the time scale on which the sequence will operate and should range over the characteristic time scale of the simulation to be tested:

\textit{Step 1.} Randomly choose an initial basis state $\ket{i}$.

\textit{Step 2.} Generate $n$ random subsets (e.g., $n=100$) of the terms of the target Hamiltonian, and define
\begin{equation}
H_{\textrm{rand},k} = \sum_{\substack{i\,\in\,\textrm{random subset}\\ \textrm{of } \{1,\,2,\,\dots,\,m\}}} H_i
\end{equation}
as the sum of the terms in subset $k$.
To increase the randomness of the resulting path, choose also the direction (forward or time-reversed) of each subset at random.
Apply each of the resulting unitary time-evolution operators, i.e.,
\begin{equation}
\label{eq:arb-step-operator}
    U_k = e^{\pm i H_{\textrm{rand},k} 2 \tau /n}
\end{equation}
for $k=1 \textrm{ to } n$, to the initial state $\ket{i}$, which evolves the system to an intermediate state $\ket{\phi}$.

\textit{Step 3.} Calculate another sequence of these random unitaries that will approximately invert the process and act on $\ket{\phi}$ to produce a basis state $\ket{f}$ within some target fidelity, e.g., 0.99.
Apply the sequence, which ideally will take the system to the final state $\ket{f}$ with probability of at least the desired target fidelity.

\textit{Step 4.} Measure the final state in the computational basis. Record the probability that the final state is measured to be $\ket{f}$.

After repeating these steps for various values of $\tau$, the resulting decay curve indicates the success probability of finding the system in the desired state after executing the randomized sequences as a function of effective simulation time.


Calculating an appropriate inversion layer, using only small time steps
of the Hamiltonian terms as building blocks, is the most computationally intensive part
of this protocol.
We cannot directly reverse the random sequence generated, since
this would simply be a time-reversal, and errors such as miscalibrations or shot-to-shot noise would cancel out.
Instead, we generate a new sequence by
explicitly calculating the product of the random sequence of unitaries and then
building a sequence which inverts it.

Since compiling an exact inversion layer (outside of simply reversing
the random sequence) is likely infeasible, we allow the inversion layer to
only approximately invert the original sequence, such that we return nearly
all of the population to a basis state. We note that the approximate nature still allows us to assess the quality of the simulation with the targeted precision using a single measurement basis. 

To construct the inversion layer,
we use the STOQ protocol for approximate compilation \cite{Shaffer2021StochasticUnitaries},
which is a stochastic Markov chain Monte Carlo (MCMC) search technique using a Metropolis-like algorithm.  This is a randomized approach to compiling an arbitrary unitary into a sequence of ``gates'' drawn from a finite set of allowed unitaries, similar to the approach used in a proposed technique for quantum-assisted quantum compiling \cite{Khatri2019Quantum-assistedCompiling}.

\begin{figure}

\includegraphics[width=1.0\linewidth]{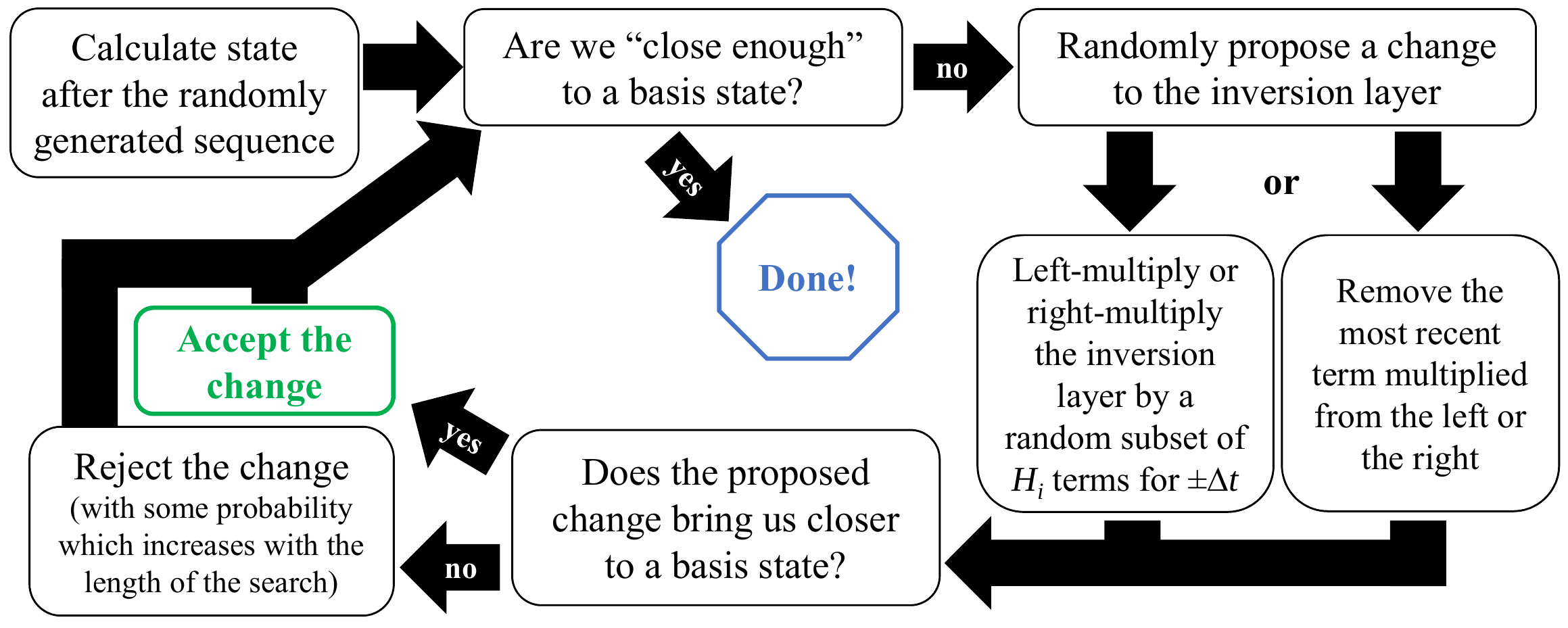}

\caption{\label{fig:approximate-compilation-flowchart}
\textbf{Illustrative flowchart for the approximate unitary compilation procedure.} The time duration $\pm\Delta t$ refers to the time and direction of each individual step in the sequence, i.e., the desired total simulation time $\tau$ divided by $n$, the total number of steps in the sequence.
}

\end{figure}

Specifically, since the set of allowed unitaries here consists of all possible random subsets of the Hamiltonian terms, we have the following procedure for approximately compiling the inversion layer (illustrated in Fig.~\ref{fig:approximate-compilation-flowchart}):
\begin{enumerate}
    \item Generate $n$ randomized layers, each of which determines a unitary operation $U_k$, as defined in Eq.~(\ref{eq:arb-step-operator}).
    \item Calculate the state after applying all $n$ of the randomized layers to the initial state as
        \begin{equation}
            \ket{\phi} = U_n U_{n-1} \cdots U_2 U_1 \ket{i} .
        \end{equation}
    \item Build up a new sequence of layers, which will become the inversion layer, by incrementally adding a randomized layer or removing a layer from the beginning or end of the sequence (such that we only have to perform one multiplication per proposed step). Let the product of these layers be $U_{\textrm{inv}}$.
    \item For each proposed addition or removal, look at the basis state of $U_{\textrm{inv}} \ket{\phi}$ with the largest population fraction to see if it has increased or decreased from the prior state. If it has increased, the system is closer to a basis state, and therefore accept the proposed addition or removal. If it has decreased, usually reject it, but sometimes accept it, based on the value of the MCMC annealing parameter $\beta$.
    \item Continue until the largest basis state population reaches some desired threshold (e.g., 0.99), which determines the population fraction in the final basis state after executing the compiled sequence.
\end{enumerate}

In order to increase the distinction between this compiled inversion sequence and the original randomly-generated sequence (which seems desirable in order to avoid potentially cancelling out any systematic errors), we initialize the MCMC search algorithm with a large value of the annealing parameter $\beta$, which increases the randomness in the early part of the compiled sequence. Over time, we linearly decrease the value of $\beta$ until the process finally converges toward a basis state.

Notice that because this procedure simply takes us approximately to some basis state (not necessarily the initial state), a true inversion sequence would require a final local rotation of the appropriate qubits to take the system back to the initial state. However, since the intention is simply to measure the resulting state, this final rotation is unnecessary -- we can just measure the state and compare the result to the expected final basis state, rather than comparing to the initial basis state.

Because this process is randomized, it is not guaranteed to converge \cite{Schkufza2013StochasticSuperoptimization}. To account for this, in the implementation used for this work, we launch many tens of MCMC search processes in parallel, which in practice typically allows the search to succeed in reasonable time. For example, in the five-qubit numerical simulation described later in this section, when the original sequence has ${\sim}100$ random layers, one of the MCMC processes will typically converge to the desired accuracy of 98\% within a few thousand steps.


The scalability of the randomized analog verification protocol is limited by
the approximate compilation of the inversion layer.
Performing this compilation requires
many explicit multiplications of unitary operators acting on the
full Hilbert space of the system being simulated, and thus has at least the same complexity as actually simulating the dynamics of the system.
Unless a reliable quantum computer is available \cite{Khatri2019Quantum-assistedCompiling}, this
must be done on a classical computer,
and so it is likely infeasible to apply this protocol
directly to systems with more than tens of qubits.

\begin{figure}

\includegraphics[width=0.5\linewidth]{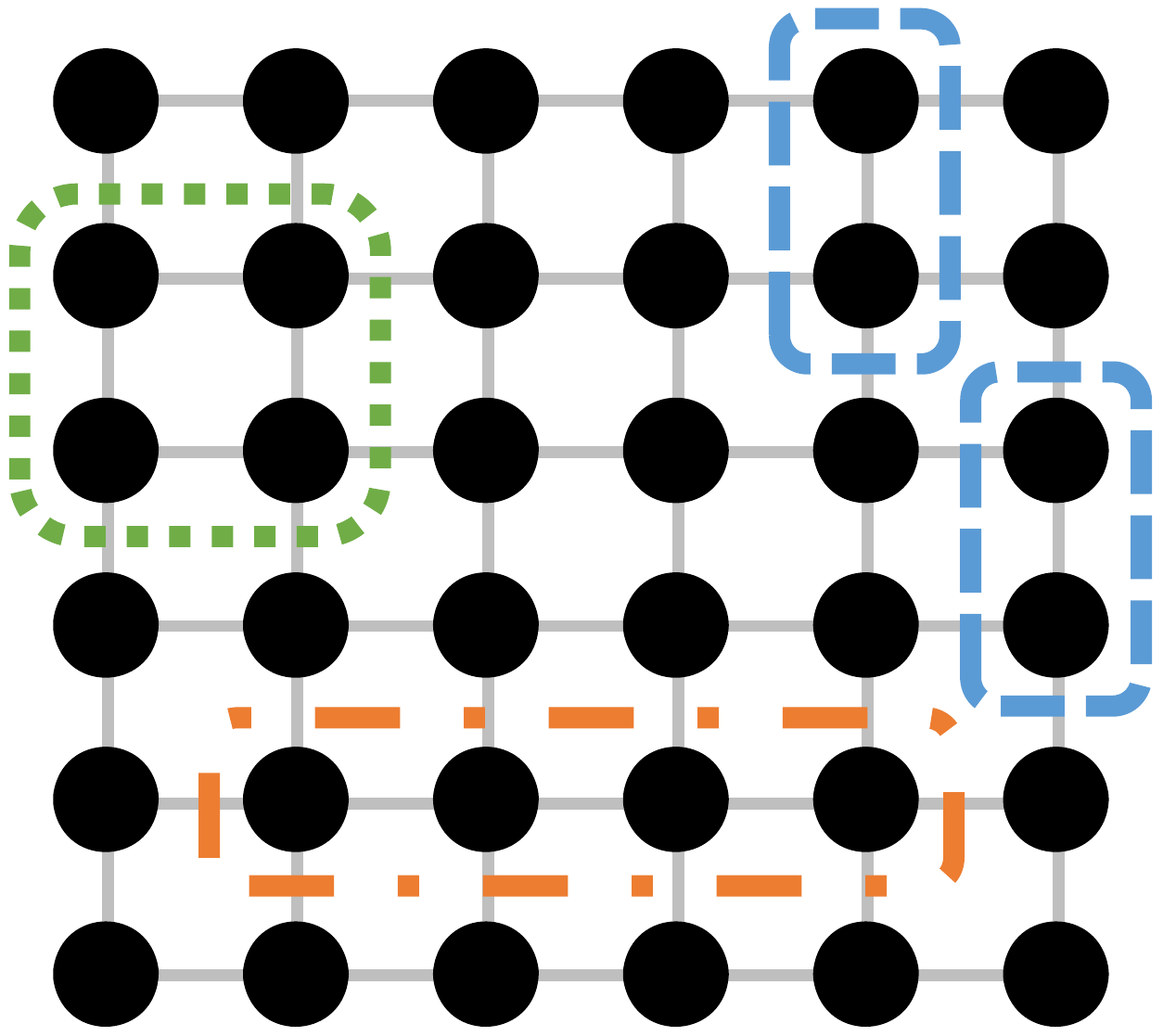}

\caption{\label{fig:subsets}
\textbf{Illustration of 6x6 2-D lattice with nearest-neighbor coupling.} Here the Hamiltonian is $k$-local with $k=2$. Colored dashed outlines show possible subsets of size $s=4$, each of which is formed from two (possibly distant) pairs of connected neighbors. The total number of ways to choose such subsets from this lattice is 3,540. Running the randomized analog verification protocol on all such subsets will test for errors associated with each interaction term in the Hamiltonian, as well as errors that may be caused by unwanted interaction (e.g., crosstalk) between any two pairs of sites in the system.
}

\end{figure}

To apply this protocol to large-scale simulations,
we can break the full system into subsystems \cite{Gambetta2012CharacterizationBenchmarking, Govia2020BootstrappingAnsatz} to reduce the exponential scaling to polynomial scaling.
Specifically, if the Hamiltonian is $k$-local, we can decompose the system
into subsystems of size $s \ge 2k$ (see Fig.~\ref{fig:subsets}), and then run this protocol on every
subsystem. This will test every interaction term, as well as
potential errors such as crosstalk that may occur between any
two distant interaction terms in the system. The number of such subsystems
grows only polynomially with degree $s$, not exponentially. Since this is equivalent to testing each subsystem of size $s$ independently, the downside of this approach is the loss of sensitivity to errors that may occur only for subsystems of size larger than $s$; however, in many systems, it is likely reasonable to assume that such errors are small. Additional work will be needed to understand exactly what claims one can make about the performance of the large-scale analog simulation by characterizing the subsystems in this way.

%
%

%
%
\subsection{\label{sec:two-site-experiment}Experimental demonstration with trapped ions}

To demonstrate the feasibility of implementing these verification protocols experimentally, we choose a simple two-site Ising model with transverse field
\begin{equation}
\label{eq:ising2q}
    H =
    - \frac{1}{2} b \left( \sigma_y^{(1)} + \sigma_y^{(2)} \right)
    - \frac{1}{2} J \sigma_x^{(1)} \sigma_x^{(2)},
\end{equation}
and we choose $J = 2\pi \times 139\textrm{ Hz}$ and $b = 2\pi \times 227\textrm{ Hz}$. We implement this model in a trapped-ion analog quantum simulator containing two $^{40}\mathrm{Ca}^+$ ions.
We use the electronic $S_{1/2}$ ground orbital and $D_{5/2}$ metastable excited orbital as the qubit states, and we drive transitions between these states using a 729 nm laser \cite{Haffner2008QuantumIons}.
In particular, we choose $\ket{g}=\ket{S_{1/2}, m_j=-1/2}$ and $\ket{e}=\ket{D_{5/2}, m_j=-1/2}$ as the states of the two-level system.

We prepare the system in the state $\ket{eg}$ or $\ket{ge}$ by optically pumping the ions to the state $\ket{gg}$, using a $\pi$-pulse with a laser beam localized to a single ion to prepare the state $\ket{eg}$, and then optionally a $\pi$-pulse with a laser beam addressing both ions to prepare the state $\ket{ge}$.

We then implement the Ising model by combining three tones in a laser beam that addresses both ions equally.
In particular, we realize the transverse field
interaction via a laser tone resonant with the qubit transition frequency with Rabi frequency $\Omega_C$. This creates the desired $(b/2)(\sigma_y^{(1)} + \sigma_y^{(2)})$ interaction with $b = \Omega_C$.
In addition, we implement the site-site coupling via a M{\o}lmer-S{\o}rensen interaction \cite{Srensen1999QuantumMotion} via the axial stretch vibrational mode with 
$\omega_{\textrm{ax}} \approx 2\pi \times 1.514\textrm{ MHz}$,
where we apply two laser tones detuned from the qubit transition frequency by 
$\pm(\omega_\textrm{ax} + \delta_\textrm{MS})$,
with $\delta_\textrm{MS} = 2\pi \times 80\textrm{ kHz}$,
and where each tone has Rabi frequency $\Omega_\textrm{MS}$.
This creates an effective $(J/2)\sigma_x^{(1)} \sigma_x^{(2)}$ interaction with $J=\eta_\textrm{ax}^2 \Omega_\textrm{MS}^2 / \delta_\textrm{MS}$, where $\eta_\textrm{ax} \approx 0.08$ is the Lamb-Dicke parameter indicating the coupling of the laser beam to the axial mode of the ion crystal, and we tune $\Omega_\textrm{MS}$ to produce the desired value of the coupling strength $J$.

%
%
\begin{figure*}

\includegraphics[width=1.0\linewidth]{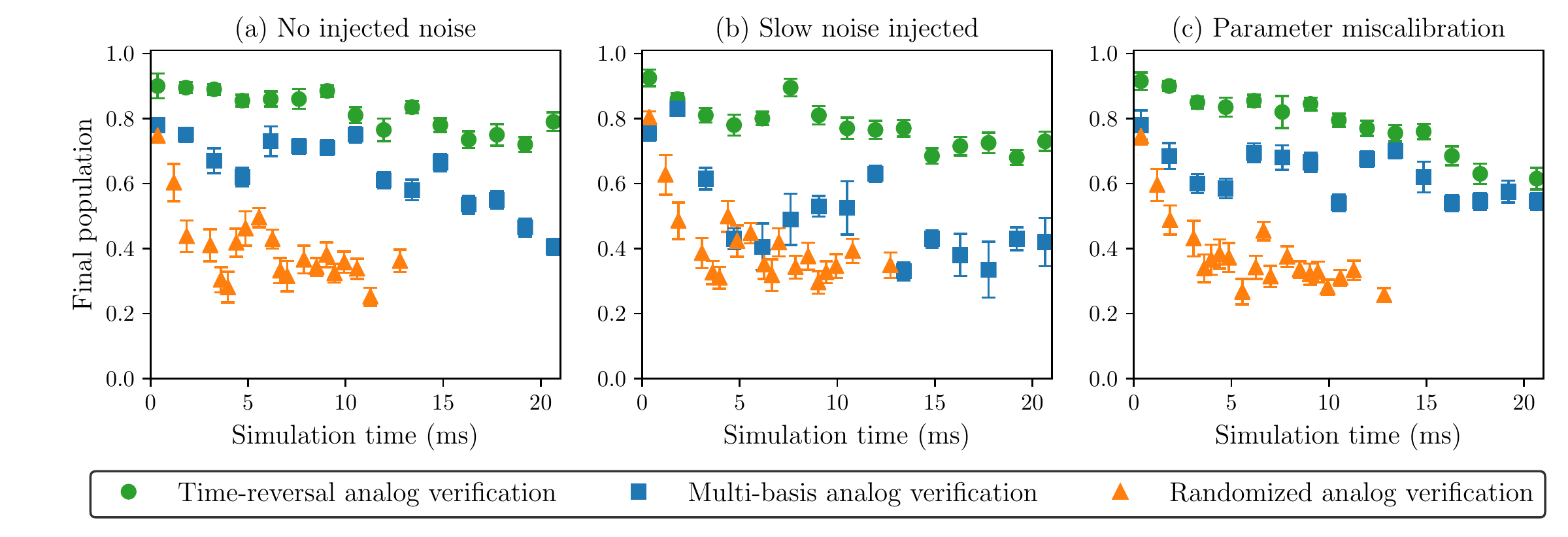}

\caption{\label{fig:experimental-plots}
\textbf{Experimental results of verification protocols.}
Results are for the two-site Ising model from Eq.~(\ref{eq:ising2q}), with $J = 2\pi \times 139\textrm{ Hz}$ and $b = 2\pi \times 227\textrm{ Hz}$.
Each plot shows the experimentally-measured population in the expected final state after running each of the verification protocols under the specified type of injected noise. Data points represent raw experimental results and include experimental errors due to state preparation, measurement, and imperfect control.
For the time-reversal and multi-basis analog verification protocols, each data point represents the distribution of measured results over 200 independent runs. For the randomized analog verification protocol, each data point represents the distribution of measured results of 10 different randomly generated sequences, with each sequence executed 100 times. Error bars indicate standard error of the mean.
}

\end{figure*}

In addition to designing the analog simulation itself, we must also implement the time-reversed and rotated versions of the simulation in order to implement the desired verification protocols. For the time-reversal analog verification protocol, we take $H$ to $-H$ by shifting the phase of the resonant tone by $\pi$, which takes $b$ to $-b$ in the transverse field interaction, and by changing the M{\o}lmer-S{\o}rensen detuning from $\delta_\textrm{MS}$ to $-\delta_\textrm{MS}$ (with a small correction to account for a change in AC Stark shift), which takes $J$ to $-J$ in the effective $\sigma_x^{(1)} \sigma_x^{(2)}$ interaction.

For the multi-basis analog verification protocol, we choose the basis rotation 
\begin{equation}
    R = R_z^{(1)}(\pi/2) + R_z^{(2)}(\pi/2),
\end{equation}
which is a global $\pi/2$ rotation around the $z$-axis. We implement $R$ physically via a sequence of single-qubit carrier rotations, using the fact that 
\begin{equation}
    R_z(\pi/2) = R_y(-\pi/2) R_x(\pi/2) R_y(\pi/2).
\end{equation}
We then must implement $RHR^\dag$, which is the Hamiltonian in the rotated basis. For the transverse field term, we note that 
\begin{equation}
    R (\sigma_y^{(1)} + \sigma_y^{(2)}) R^\dag = \sigma_x^{(1)} + \sigma_x^{(2)},
\end{equation}
which we implement by simply shifting the phase of the resonant tone by $\pi/2$ as compared to the phase used to implement $\sigma_y^{(1)} + \sigma_y^{(2)}$. For the coupling term, we note that 
\begin{equation}
    R \sigma_x^{(1)} \sigma_x^{(2)} R^\dag = \sigma_y^{(1)} \sigma_y^{(2)},
\end{equation}
which we implement by shifting the phase of the blue-sideband M{\o}lmer-S{\o}rensen tone by $\pi$ with respect to the red-sideband tone \cite{Lee2005PhaseGates}.

Finally, for the randomized analog verification protocol, we write the target Hamiltonian from Eq.~(\ref{eq:ising2q}) as
\begin{equation}
    H=H_1+H_2,
\end{equation}
where $H_1$ and $H_2$ are defined as
\begin{align}
    H_1 &= - \frac{1}{2} b ( \sigma_y^{(1)} + \sigma_y^{(2)} ), \\
    H_2 &= - \frac{1}{2} J \sigma_x^{(1)} \sigma_x^{(2)}.
\end{align}
We then generate 200 random sequences of subsets of these Hamiltonian terms in either the forward or time-reversed direction, such that each step of each sequence is selected from the set
\begin{equation}
    H_\textrm{steps} = \{ H_1, H_2, H_1 + H_2, -H_1, -H_2, -H_1-H_2 \},
\end{equation}
and each sequence consists of $10 \le n \le 50$ steps of length $8\ \mu\textrm{s} \le t_\textrm{step} \le 290\ \mu\textrm{s}$.
For each sequence, we then compile an approximate inversion sequence consisting of steps from the same set $H_\textrm{steps}$. Each sequence has a randomly-chosen initial state from the set $\{\ket{ge}, \ket{eg}\}$, and each full sequence ideally leaves the system in some basis state with at least 98\% probability.
The terms in the set $H_\textrm{steps}$ are implemented experimentally by enabling or disabling the corresponding laser tones and by time-reversing the analog simulation as necessary.

To test the behavior of each of these protocols, we execute the time-reversal and multi-basis analog verification protocols for varying simulation times and execute all 200 of the randomized analog verification sequences. The results of these experimental runs are shown in Fig.~\ref{fig:experimental-plots}. To produce these results, we executed each protocol under three different sets of experimentally-motivated noise conditions:
\begin{enumerate}
    \item \textit{No injected noise:} We execute each of the verification protocols after calibrating the individual interactions to approximately match the desired dynamics.
    \item \textit{Slow noise injected:} We introduce shot-to-shot fluctuations by intentionally varying the intensity of each of the three tones in the laser beam using parameters drawn from a Gaussian distribution with relative standard deviation of 3~dB. The parameter variations in the original basis are drawn independently from those in the rotated basis, which emulates the case where the system has independent noise sources in the two bases.
    \item \textit{Parameter miscalibration:} We intentionally miscalibrate the M{\o}lmer-S{\o}rensen detuning to $\delta_\textrm{MS} = 2\pi \times 60\textrm{ kHz}$, which has the effect of increasing the coupling strength $J$ by a factor of $1/3$.
\end{enumerate}

\begin{figure}
    \centering
    \includegraphics[width=0.9\linewidth]{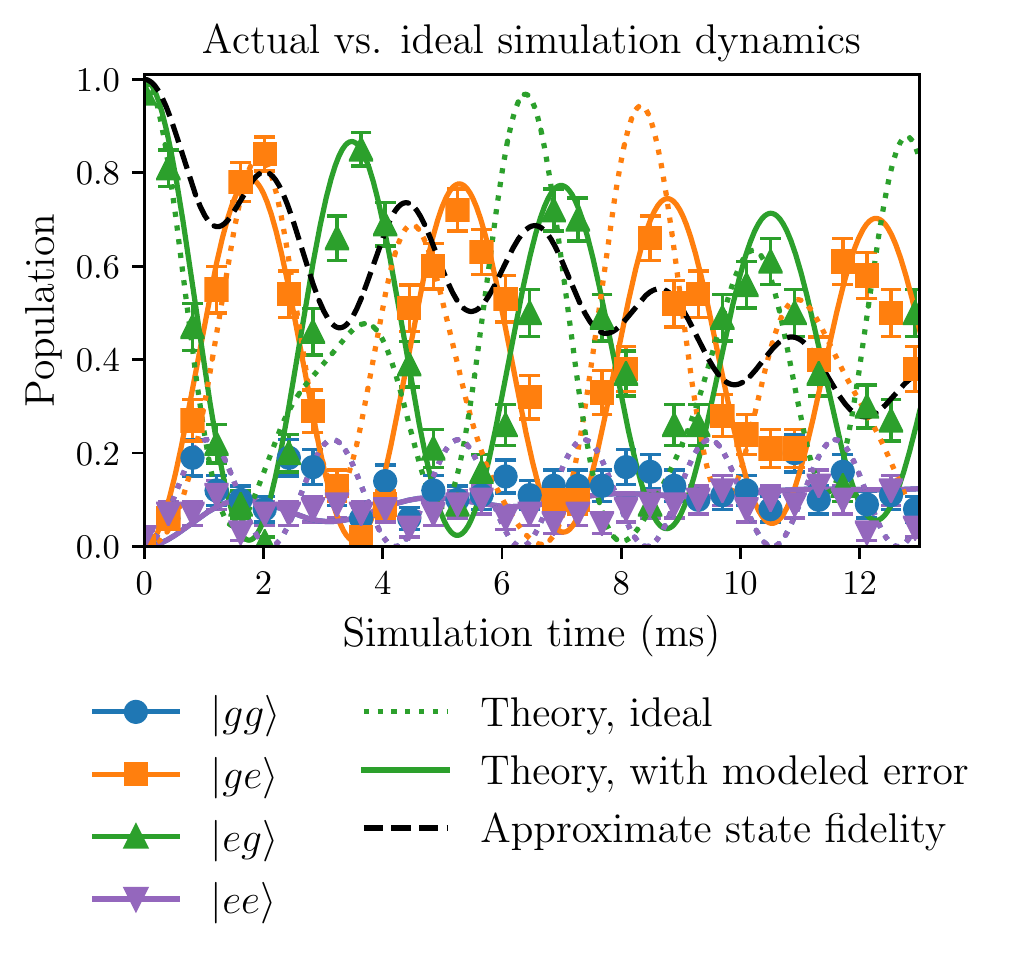}
    \caption{
    \label{fig:experimental-dynamics}
    \textbf{Experimentally-measured dynamics of the simulation as a function of time.}
    Each data point is the average of 100 independent runs. Error bars indicate standard error of the mean. The dotted curves represent the ideal dynamics of a perfectly-calibrated analog simulation ($J=2\pi \times 139$ Hz, $b=2\pi \times 227$ Hz) in the absence of noise. The solid curves represent the theoretical dynamics of a miscalibrated analog simulation ($J=2\pi \times 250$ Hz, $b=2\pi \times 102$ Hz) with a dephasing rate of $\gamma_\phi = 2\pi \times 38$ Hz, where these parameters are chosen empirically as a reasonable approximation of the observed experimental data points.
    The dashed curve is the fidelity of the state evolved according to the miscalibrated dynamics with the state evolved according to the ideal dynamics, calculated using Eq.~(\ref{eq:fidelity}).
    }
\end{figure}

To provide more insight into the results of these protocols, in Fig.~\ref{fig:experimental-dynamics} we plot the actual population dynamics of the analog simulation in the absence of injected noise. We observe that the implemented simulation diverges significantly from the ideal simulation after only a few milliseconds, primarily due to miscalibration and dephasing noise. We intentionally allow this divergence as a test case for the various verification protocols, since it is caused by errors that may be typical in experiments. The miscalibration here is due to laser intensities and/or frequencies that have not been optimized to produce the desired dynamics, and the dephasing noise is likely caused by the presence of global magnetic field fluctuations which cause the state to decohere when leaving the subspace $\{\ket{ge}, \ket{eg}\}$, which is a decoherence-free subspace with respect to the global magnetic field.

Also plotted in Fig.~\ref{fig:experimental-dynamics} is a curve showing the fidelity between an ideal evolution of the system state and an approximation of the system state obtained experimentally.
For the ideal Hamiltonian $H$, defined in Eq.~(\ref{eq:ising2q}), we use the target values ($J=2\pi \times 139$ Hz, $b=2\pi \times 227$ Hz) and perform unitary evolution under the Schr{\"o}dinger equation to obtain the dynamics of the ideal state $\rho(t) = \ket{\psi(t)} \bra{\psi(t)}$, where $\ket{\psi(t)} = e^{-i H t} \ket{\psi(0)}$. For the experimentally-miscalibrated Hamiltonian $\tilde{H}$, we use parameters that approximately match the observed measurements ($J=2\pi \times 250$ Hz, $b=2\pi \times 102$ Hz) with an appropriate dephasing rate ($\gamma_\phi = 2\pi \times 38$ Hz). We then perform non-unitary evolution under the Lindblad master equation, using the Lindblad operator $L = \sqrt{\gamma_\phi/2}\, \sigma_z$ as the dephasing mechanism, to obtain the approximate dynamics of the experimentally-obtained state $\tilde{\rho}(t)$. The approximate fidelity between the ideal state and the experimentally-obtained state is then
\begin{equation}
\label{eq:fidelity}
    \tilde{F}(t) = \left[ \tr \sqrt{ \sqrt{\rho(t)}\, \tilde{\rho}(t)\, \sqrt{\rho(t)} } \right]^2.
\end{equation}
The fidelity curve plotted in Fig.~\ref{fig:experimental-dynamics} is this approximate fidelity function $\tilde{F}(t)$, and we observe that it decays to 50\% in approximately 7 ms.

Despite this fast decay of the fidelity, we note that in the absence of additional injected noise, both the time-reversal and multi-basis analog verification protocols in Fig.~\ref{fig:experimental-plots}(a) show decay times on the order of tens of milliseconds.
Because these protocols are sensitive to fast, incoherent noise, we deduce that the majority of the errors present in the experiment are slower than the timescale of each experiment and are therefore cancelled out by these protocols.

Conversely, we consider the results of the randomized analog verification protocol with no injected noise in Fig.~\ref{fig:experimental-plots}(a). The success probability decays in approximately 3 ms, which is slightly faster than the fidelity decay observed in Fig.~\ref{fig:experimental-dynamics}.
This suggests that the randomized protocol at least detects these experimental miscalibrations or coherent errors that cause the actual simulation dynamics to differ from the ideal dynamics.
That is, the randomized analog verification protocol helps to identify imperfections in the simulation with respect to the target Hamiltonian, which is something that the other protocols are unable to do.
In addition, the faster decay of the randomized analog verification results as compared to the approximate fidelity curve in Fig.~\ref{fig:experimental-dynamics} indicates that there are additional sources of experimental error
that are not captured by the population dynamics alone.
For example, the experimental procedure involves rapidly enabling and disabling the various interaction terms, which may itself introduce imperfections that cause the success probability to decay more rapidly.
Indeed, the difference between the randomized analog verification protocol results with no injected noise in Fig.~\ref{fig:experimental-plots}(a) and with injected noise in Fig.~\ref{fig:experimental-plots}(b) and Fig.~\ref{fig:experimental-plots}(c) indicate that the experimental errors in the simulation dwarf the errors caused by the injected noise.

Finally, we note that a number of the experimental data series in Fig.~\ref{fig:experimental-plots} show hints of oscillatory behavior, and that in general the shape of each decay curve is non-exponential. This is evidence supporting the claim that these protocols do not fully twirl coherent errors into incoherent errors, and thus do not produce a fully depolarizing channel that would produce an exponential decay in these results.

%
%

%
%
\begin{figure*}
\includegraphics[width=1.0\linewidth]{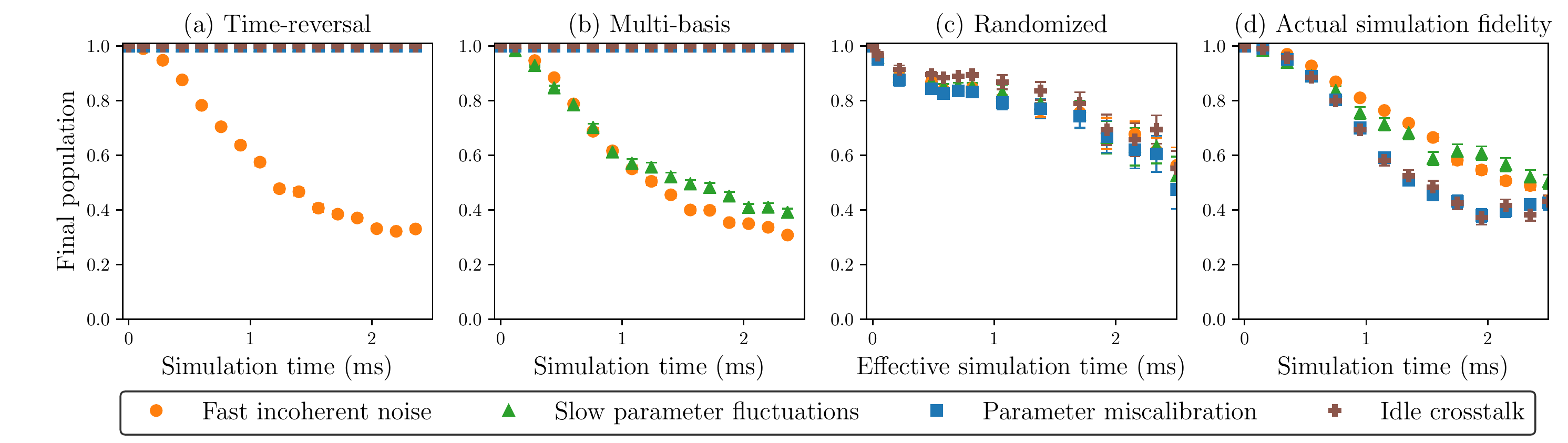}
\caption{\label{fig:numerical-plots}
    \textbf{Numerical five-qubit simulation results of verification protocols under simulated noise conditions.}
Five-qubit numerical simulation results showing the sensitivities of each of the three analog verification protocols to four different types of experimental error sources:
fast incoherent noise (${\sim}\tau/n$) in Hamiltonian parameters with 30\% relative standard deviation (RSD),
slow fluctuations (${\gg}\tau$) in Hamiltonian parameters with 15\% RSD,
constant miscalibration of Hamiltonian parameters with 10\% RSD,
and constant idle crosstalk affecting all sites with 10\% RSD.
The target Hamiltonian is the five-qubit Heisenberg model from Eq.~(\ref{eq:heisenberg5q}), with $b^{(i)} = J_x^{(i,j)} = J_y^{(i,j)} = J_z^{(i,j)} = 2\pi \times 1\textrm{ kHz}$. The ``effective simulation time'' is the average time for which each term of the Hamiltonian is enabled. Error bars indicate standard error of the mean. \\
(a) Time-reversal analog verification results. Each data point represents the distribution of results over 50 runs. \\
(b) Multi-basis analog verification results. Each data point represents the distribution of results over 50 runs. \\
(c) Randomized analog verification results. Each data point represents the distribution of 10 different randomly generated sequences with $n=150$ steps, with each sequence simulated 20 times. \\
(d) Actual fidelity of the analog simulation under each type of noise.
}

\end{figure*}

%
%
\subsection{\label{sec:five-site-simulation}Numerical demonstration under simulated noise conditions}

To further test the sensitivity of each protocol to various types of noise, we numerically simulated the dynamics of the verification protocols using the five-site Heisenberg model
\begin{equation}
\label{eq:heisenberg5q}
\begin{split}
    H =
    - \frac{1}{2} \sum_{i=1}^{5} b^{(i)} \sigma_z^{(i)}
    - \frac{1}{2} \sum_{i=1}^{4}
        \Big(
            J_x^{(i,i+1)} \sigma_x^{(i)} \sigma_x^{(i+1)} + \\
            J_y^{(i,i+1)} \sigma_y^{(i)} \sigma_y^{(i+1)} +
            J_z^{(i,i+1)} \sigma_z^{(i)} \sigma_z^{(i+1)} \Big).
\end{split}
\end{equation}
Nominally, we fix all parameter values as $b^{(i)} = J_x^{(i,j)} = J_y^{(i,j)} = J_z^{(i,j)} = 2\pi \times 1\textrm{ kHz}$, but we vary each of these parameters during the simulation according to several different types of potential experimental noise.
We simulated the dynamics of each protocol under conditions with
several classes of noise sources present individually:
\begin{enumerate}
    \item \textit{Fast incoherent noise:} The $b$ and $J$ terms in the Hamiltonian have fast noise, modeled as an Ornstein-Uhlenbeck process with a correlation time on the order of $\tau/n$, which is approximately the duration of one step of the randomized analog verification protocol.
    \item \textit{Slow parameter fluctuations:} The $b$ and $J$ terms in the Hamiltonian have slow noise (modeled as a constant miscalibration that varies from run to run with a Gaussian distribution) that has a typical timescale longer than $\tau$, but shorter than the time between individual experiments.
    \item \textit{Parameter miscalibration:} Each of the $b$ and $J$ terms in the Hamiltonian is miscalibrated from the desired value.
    \item \textit{Idle crosstalk:} Each of the interaction terms in the Hamiltonian, when disabled, still drives the interaction with some fraction of the intended strength. For example, during steps of the randomized analog verification protocol in which the $\sigma_y^{(1)} \sigma_y^{(2)}$ interaction is intended to be turned off, we still include a fraction of that term in the Hamiltonian being simulated.
\end{enumerate}

The numerical simulation results in Fig.~\ref{fig:numerical-plots} demonstrate that certain types of noise, such as fast incoherent noise, can be detected by any of the proposed verification protocols. We see that the multi-basis analog verification protocol is also sensitive to certain slow parameter fluctuations, whereas the randomized analog verification protocol is additionally sensitive to errors such as parameter miscalibration
and crosstalk among the interaction terms in the system.
Such error sources may cancel out in the forward and backward directions when using more systematic protocols \cite{Ball2016EffectBenchmarking,Edmunds2017MeasuringCircuits}, but when using a randomized protocol they are highly unlikely to cancel due to the randomized nature of the sequence and its dependence on the exact parameters of the Hamiltonian. In particular, we see in Fig.~\ref{fig:numerical-plots}(d) that the actual fidelity of the analog simulation is most severely impacted by the parameter miscalibration and crosstalk errors, and only the randomized analog verification protocol is able to detect the presence of these errors.

%
%
\begin{figure*}
\includegraphics[width=1.0\linewidth]{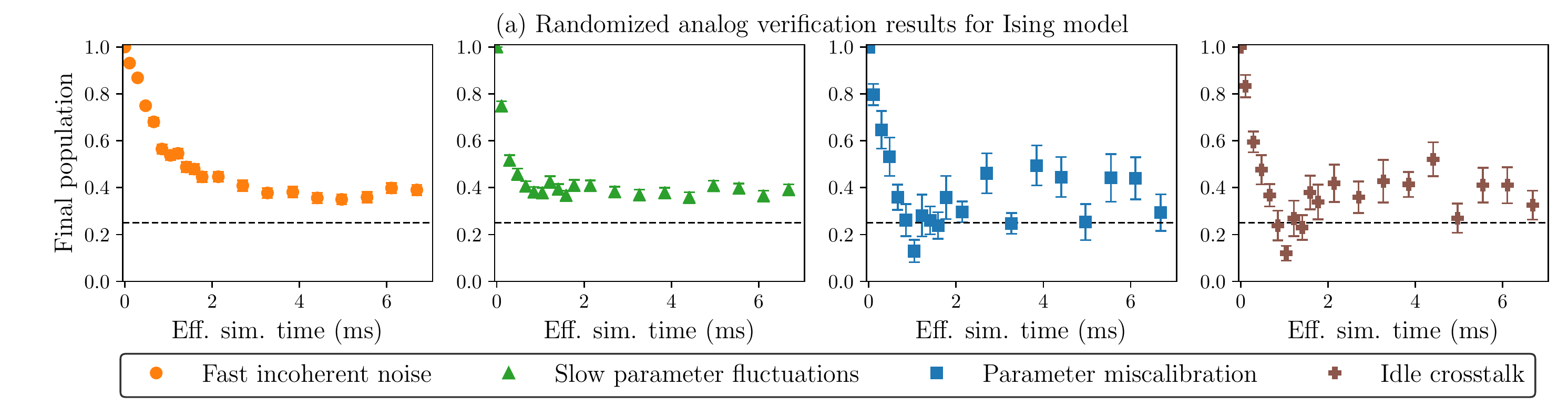}
\includegraphics[width=1.0\linewidth]{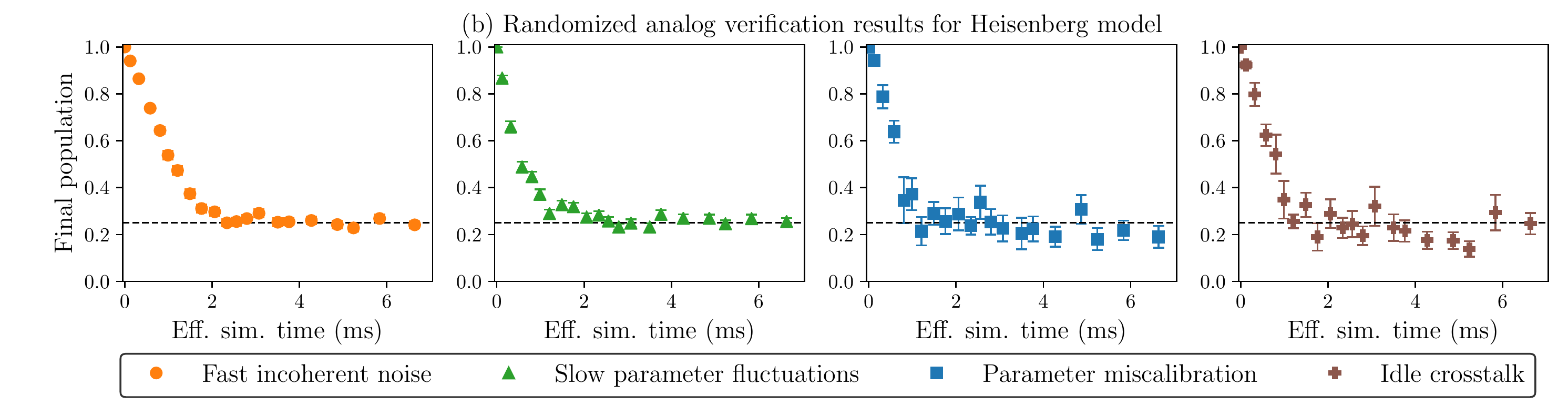}
\caption{\label{fig:two-qubit-numerical-plots}
    \textbf{Numerical two-qubit simulation results of verification protocols for two different Hamiltonians.}
Two-qubit numerical simulations showing randomized analog verification results for two Hamiltonians under four different types of experimental error sources:
fast incoherent noise (${\sim}\tau/n$) in Hamiltonian parameters,
slow fluctuations (${\gg}\tau$) in Hamiltonian parameters,
constant miscalibration of Hamiltonian parameters,
and constant idle crosstalk affecting all sites.
 The ``effective simulation time'' is the average time for which each term of the Hamiltonian is enabled.
Each data point represents the distribution of 10 different randomly generated sequences, with each sequence simulated 15 times. Error bars indicate standard error of the mean. Dashed line added at $y=0.25$ on each plot as a visual aid. \\
(a) The target Hamiltonian is the one-dimensional Ising model from Eq.~(\ref{eq:rav-ising2q}) under fast incoherent noise with 12\% relative standard deviation (RSD), slow parameter fluctuations with 6\% RSD, parameter miscalibration with 3\% RSD, and idle crosstalk with 3\% RSD.  \\
(b) The target Hamiltonian is the one-dimensional Heisenberg model from Eq.~(\ref{eq:rav-heisenberg2q}) under fast incoherent noise with 4\% RSD, slow parameter fluctuations with 2\% RSD, parameter miscalibration with 1\% RSD, and idle crosstalk with 1\% RSD. \\
For both (a) and (b), we choose $b = J_x = J_y = J_z = 2\pi \times 20\textrm{ kHz}$.
Note that larger relative errors are used in (a) to compensate for the smaller number of Hamiltonian terms in this simulation, such that the decay times of the plots in (a) and (b) are similar.
}

\end{figure*}


To gain further insight into the behavior of the randomized analog verification protocol, we also simulated the dynamics under various types of noise using a pair of two-qubit Hamiltonians. First, we use a one-dimensional Ising model with transverse field
\begin{equation}
\label{eq:rav-ising2q}
    H =
    - \frac{1}{2} \Big(
        b (\sigma_y^{(1)} + \sigma_y^{(2)}) +
            J_x \sigma_x^{(1)} \sigma_x^{(2)} \Big),
\end{equation}
which is identical to Eq.~(\ref{eq:ising2q}), the Hamiltonian used for the experiment.
For the purposes of the randomized analog verification protocol, we treat $b (\sigma_y^{(1)} + \sigma_y^{(2)})$ as a single term, as was also done in the experiment.

Second, we use a one-dimensional Heisenberg model with transverse field terms along each axis
\begin{equation}
\label{eq:rav-heisenberg2q}
\begin{split}
H =
    - \frac{1}{2} \Big(
        b \sigma_x^{(1)} +
            b \sigma_y^{(1)} +
            b \sigma_z^{(1)} + \qquad\qquad\qquad\quad \\
        b \sigma_x^{(2)} +
            b \sigma_y^{(2)} +
            b \sigma_z^{(2)} + \qquad\qquad\qquad\quad \\
            J_x \sigma_x^{(1)} \sigma_x^{(2)} +
            J_y \sigma_y^{(1)} \sigma_y^{(2)} +
            J_z \sigma_z^{(1)} \sigma_z^{(2)} \Big),
\end{split}
\end{equation}
which is a simplified version of the five-qubit Hamiltonian in Eq.~(\ref{eq:heisenberg5q}) used for the earlier simulations.

Fig.~\ref{fig:two-qubit-numerical-plots} contains the numerical simulation results of applying the randomized analog verification protocol to these two Hamiltonians under various types of noise, where we have chosen $b = J_x = J_y = J_z = 2\pi  \times 20\textrm{ kHz}$ such that the effective simulation times are much longer than the timescale of the system dynamics.

We note that the shape of the decay differs significantly between the two Hamiltonians. In particular, we observe that each of the decay curves for the Heisenberg model in Fig.~\ref{fig:two-qubit-numerical-plots}(b) appears to be nearly exponential in shape and decays to approximately 0.25, which is the expected result for a fully mixed two-qubit state. This is not the case for some of the decay curves for the Ising model in Fig.~\ref{fig:two-qubit-numerical-plots}(a).

As discussed previously, randomized benchmarking protocols produce exponential decay curves in cases where the noise is fully depolarized by the randomized circuits. We note that the ``native gate set'' obtained from the Heisenberg model in Eq.~(\ref{eq:rav-heisenberg2q}) is a universal set of quantum gates, which forms an approximate 2-design in the limit of long sequence length. Here we are in fact operating in the limit of ``long sequence length'', since the dynamics occur at 20 kHz and the protocol is being performed for an effective simulation time of a few milliseconds. So the nearly-exponential shape of the decay curves in Fig.~\ref{fig:two-qubit-numerical-plots}(b) is a good indication that the various noise sources are indeed being depolarized under these conditions.

In contrast, the behavior of the decay curves in Fig.~\ref{fig:two-qubit-numerical-plots}(a), which do not decay to 0.25, can be explained by the fact that the interactions do not fully explore the state space of the system. We also observe non-monotonic behavior of these decay curves in the presence of correlated errors such as miscalibration or crosstalk, which suggests that such errors are not being fully depolarized. Such non-monotonic behavior is also observed in the experimental data in Fig.~\ref{fig:experimental-plots}.

%
%
\section{\label{sec:discussion}Discussion}

The set of verification protocols for analog quantum simulators introduced in this work are experimentally motivated, and we have demonstrated the utility of these protocols both experimentally and numerically. Most notably, we observe that the randomized analog verification protocol is superior in terms of the types of experimental errors to which it is sensitive, but that its scalability to large system sizes requires additional assumptions, such as the ability to verify subsets of the system independently, due to the classical resources required to perform the approximate inverse compilation during the generation of the randomized sequences. We also observe that the randomized analog verification protocol produces results similar to those from traditional randomized benchmarking protocols in cases where the Hamiltonian terms form a universal ``native gate set'' and where the simulation time is long in comparison to the system dynamics.

It is worth noting that implementing the time-reversed Hamiltonian in the analog quantum simulation device, which is required for all of the discussed verification protocols, is not necessarily trivial for general Hamiltonians that may be simulated. It turns out that the slow M{\o}lmer-S{\o}rensen interaction used to implement the Ising model with trapped ions is easily time-reversible, as demonstrated in the experimental results, which allowed us to demonstrate each of the protocols here without much additional effort. It is likely that many interactions of interest on other physical platforms, such as neutral atoms or superconducting qubits, may have similarly simple physical mechanisms for time-reversing the dynamics.

Ideally, verification protocols are useful not only for verifying the correct behavior of a system, but also for helping to diagnose and fix errors.
In particular, an experimentalist may wish to identify not only the existence of errors in the system, but also the types and locations of these errors.
Simply running the dynamics of the full simulation and checking the results may not provide the information necessary to diagnose these details.
However, the protocols described in this work provide additional tools for the experimentalist to help characterize the errors in the system.
For example, running each of the protocols and comparing the relative decay curves could help to provide insight into whether the system suffers from fast incoherent noise, slow parameter fluctuations, parameter miscalibration, and/or crosstalk errors.
In addition, because each of the protocols can be run on arbitrary subsets of the full system, running each on many different subsets will help to isolate the problematic physical interactions.

One may also consider whether such protocols could have application in the validation of gate-based quantum computers. At present, and until error-corrected devices become a reality, quantum computers are realized by carefully tuning the underlying analog interactions to implement quantum gates with the highest fidelity possible. Because the underlying interactions are analog, these analog verification protocols could be adapted for use in verifying the behavior of gate-based devices as well. The randomized analog verification protocol may be practically scalable to larger numbers of qubits than traditional RB because it directly uses the native interactions of the device, rather than requiring compilation of arbitrary Clifford gates into native gates. Most scalable variants of RB, such as direct RB \cite{Proctor2019DirectDevices}, require that the native gate set be a generator of the Clifford group, or at least that the native gate set is universal. Randomized analog verification imposes no restriction on the types of interactions present in the Hamiltonian, since it does not rely on properties of the gate set to efficiently return the state to the measurement basis. Of course, this is also a limitation of the randomized analog verification protocol, since one cannot make strong claims on the shape or meaning of the resulting decay curve without limitations on the gate set.

An important feature of the randomized analog verification protocol is the efficiency of executing the experiments on the physical device. Because the protocol requires measurement only in a single basis, the number of measurements required is not only significantly fewer than performing full tomography, but it also significantly fewer than sampling-based techniques such as cross-entropy benchmarking \cite{Boixo2018CharacterizingDevices}. This is enhanced by the fact that the protocol measures the system when it is near a measurement basis state, which minimizes the effect of quantum projection noise \cite{Itano1993QuantumSystems, Monz2011QuantumIon-qubits} and therefore reduces the number of measurements required in order to achieve a desired level of accuracy in the fidelity estimate.
To check that the system is in a particular basis state, only 100 repetitions would be required to verify a fidelity of 99\% to within 1\% error. But for an arbitrary state, the projection noise scales as the inverse square root of the number of measurements, which would require on the order of 10,000 measurements to achieve a similar level of precision.

%
%

This work has introduced three experimentally-motivated verification protocols for validation of analog quantum simulators and has demonstrated the feasibility of these protocols both numerically and experimentally. Taken together, these techniques allow for pragmatic evaluation of an analog quantum simulation device in a way that builds confidence that the device is not only operating consistently, but that it is also operating faithfully according to the desired target Hamiltonian. The decay curves resulting from these protocols may then provide some insight into the type and strength of errors encountered. Such techniques can also be applied to subsets of a larger system to allow an experimentalist to characterize and diagnose the behavior in a scalable way. Future work should pursue a more detailed analysis of the information that these protocols can provide about the types of noise or errors present in the system, the feasibility of applying the randomized analog verification protocol to larger systems, and the application of similar techniques to gate-based quantum computing devices. In addition, alternative protocols should be explored that combine the ideas in these protocols with existing techniques from the randomized benchmarking literature, with the goal of producing a practical protocol which depolarizes errors more fully and about which stronger theoretical claims can be made with regard to noise sensitivity and the expected shape of the decay curve.

%
%


%
%
\section*{Acknowledgments}

We thank Clemens Matthiesen, Sara Mouradian, Mohan Sarovar, and Birgitta Whaley for feedback on early drafts of this manuscript, and we thank Sumanta Khan for work on the experimental apparatus.
R.S. acknowledges government support under contract FA9550-11-C-0028 and awarded by the Department of Defense, Air Force Office of Scientific Research, National Defense Science and Engineering Graduate (NDSEG) Fellowship, 32 CFR 168a.
Work by E.M. and development of the experimental system were supported by the Army Research Office under Grant Number W911NF-18-1-0170.
J.B. acknowledges support via the NSF Graduate Research Fellowships Program.
Work by W.C. and H.H. was sponsored by the U.S. Department of Energy, Office of Science, Office of Basic Energy Sciences under Award Number DE{-}SC0019376.
The views and conclusions contained in this document are those of the authors and should not be interpreted as representing the official policies, either expressed or implied, of the Army Research Office or the U.S. Government. The U.S. Government is authorized to reproduce and distribute reprints for Government purposes notwithstanding any copyright notation herein. This is a post-peer-review, pre-copyedit version of an article published in \textit{npj Quantum Information}. The final authenticated version is available online at: \url{http://dx.doi.org/10.1038/s41534-021-00380-8}.

%
%
\section*{Author Contributions}

R.S. developed the verification protocols, performed the numerical simulations and experimental data analysis, and wrote the manuscript. R.S., E.M., and H.H. developed the main ideas and contributed to the interpretation of the results. R.S., J.B., and W.C. contributed to the design and execution of the experiments. All authors contributed to discussions of the results and provided revisions to the manuscript.

%
%


%
%
\bibliography{references-npjqi}

%
%

\end{document}